\documentstyle[12pt,epsfig]{article}
\newcommand{\be}{\begin{equation}}
\newcommand{\ee}{\end{equation}}
\newcommand{\Tr}{\mathop{\rm Tr}\nolimits}

\begin{document}
\title{
\begin{flushright}
{\small SMI-25-97 }
\end{flushright}
\vspace{2cm}
Chaos in M(atrix) Theory}

\author{
I. Ya. Aref'eva${}^{\S}$, P. B. Medvedev${}^{\star}$,
O. A. Rytchkov${}^{\dag}$,\\ and \\ I. V. Volovich${}^{\S}$\\
\\${}^{\S}$ {\it  Steklov Mathematical Institute,}\\ {\it Gubkin st.8, Moscow,
Russia, 117966}\\
arefeva,volovich@genesis.mi.ras.ru\\
\\${}^{\star}$
{\it Institute of Theoretical and Experimental Physics,}\\
{\it B.Cheremushkinskaya st.25, Moscow, 117218}\\
medvedev@heron.itep.ru\\
\\${}^{\dag}$
{\it Physical Department, Moscow State University, }\\
{\it Moscow, Russia, 119899} \\
rytchkov@grg1.phys.msu.su
}

\date {$~$}
\maketitle
\begin {abstract}
We consider the classical and quantum dynamics in M(atrix) theory. Using a
simple ansatz we show that a classical trajectory exhibits a chaotic motion.
We argue that the holographic feature of M(atrix) theory is related with the
repulsive feature of energy eigenvalues in quantum chaotic system.  Chaotic
dynamics  in $N=2$ supersymmetric Yang-Mills theory is also discussed.
We demonstrate that after the separation of "slow"
and "fast" modes there is a singular contribution from the
"slow"  modes to the Hamiltonian of the "fast" modes.
\end {abstract}

\newpage
Recent advances in string theory have led to the discovery of dualities
between  five known superstring theories. These theories are expected to
be obtained by taking various limits of
the conjectured eleven-dimensional M-theory \cite {HT,Wit}.
At low energies/large distances M-theory
is described by eleven-dimensional supergravity.
Banks, Fischler, Shenker and Susskind
\cite {BFSS} have proposed that M-theory in the infinite momentum frame
is described in terms of a supersymmetric matrix model,
the so called M(artix) theory. Moreover,
the only dynamical degrees of freedom are Dirichlet zero-branes
and the calculation of any physical quantity of M-theory can
be reduced to a calculation in the matrix quantum mechanics.
A system of $N$ Dirichlet zero-branes is described in terms of
nine $N\times N$ Hermitian matrices $X_i, i=1,...,9$ together
with their fermionic superpartners. The action can be regarded
as ten-dimensional $SU(N)$ supersymmetric Yang-Mills theory
reduced to $(0+1)$ space-time dimensions:

\begin{equation} \label {1.1}
S=\int dt Tr(\frac{1}{2}D_tX_iD_tX_i+\frac{1}{4}
[X_i,X_j][X_i,X_j] )+ (fermions),
\end{equation}
where $D_t=\partial_t +iA_0$.
The action (\ref {1.1}) was considered in
the theory of eleven-dimensional
supermembranes in \cite {WHN,WLN,FH} and in the dynamics of D-particles
in  \cite {DFS,KP}. In the original formulation \cite {BFSS}
of the conjectured correspondence between M-theory and M(atrix) theory
the large $N$ limit was assumed. A more recent formulation
\cite {Sus} is valid for finite $N$. The M(atrix) theory is interesting
because it could provide a non-perturbative approach to quantum gravity.
Therefore it is important to investigate exact properties of the model
(\ref {1.1}).

In this paper we study the bosonic part of the dynamical
system (\ref {1.1}). We show a complicated chaotic behavior
 of classical trajectories and discuss its quantization.
The appearance of chaos in a
classical system means that we cannot trust to
the ordinary perturbative analysis of the corresponding
quantum system. We demonstrate that after the separation of "slow"
and "fast" modes there is a singular contribution from the
"slow"  modes to the Hamiltonian of the "fast" modes (see formulae
(\ref {h1}) and (\ref {h2})).
We also discuss a possibility to consider quantum chaos as a source of the
holographic feature of the M(atrix) theory.

A classical dynamical system is defined by
its phase space $P$, the dynamical flow $S_t$
and the invariant measure $\mu$. For the Hamiltonian
system one considers  the reduced phase space obtained by fixing
the energy and other (if any) integrals  of motion. The measure
$\mu$ is the corresponding restriction of the Liouville measure
$\prod dq_idp_i$. The system is said to exhibit a chaotic or stochastic
behavior if it is ergodic and moreover it is unstable, i.e. it has a
positive Lyapunov exponent.

There are different levels of chaos and they
can be specified for instance by  $K$-property, central limit theorem,
exponential decay of correlations etc. \cite {AA}-\cite {Ohya}.
Chaos is often quantified by
computing Lyapunov exponents.
If the dynamical system is defined by means of the system of differential
equations
$\dot {x}_{i}=F_{i}(x)$ then the Lyapunov exponent $\chi$ of a solution
$x_{i}(t)$ is given by
\begin{equation} \label {1.9}
\chi =\lim _{t \to \infty} \frac{1}{t}
\log \frac{\rho (t)}{\rho (0)},
\end{equation}
where
$$
\rho^2 (t)=\sum _{i}(a_{i}^{2}(t)+\dot{a}_{i}^{2}(t))
$$
and  $a_{i}(t)$ is the solution of the equation
$\dot{a}=F^{'}(x)a$.

The equations  of motion for the action (\ref {1.1}) in the $A_0=0$
gauge read
\begin {equation} \label {1.2}
\ddot{X}_i =[[X_j,X_i],X_j].
\end{equation}
By varying over $A_0$ one  also gets the constraint
\begin {equation} \label {1.3}
[X_i,\dot{X}_i]=0.
\end{equation}

The Hamiltonian  for the bosonic part of the action (\ref {1.1}) reads
\begin{equation} \label {1.4}
H=Tr(\frac{1}{2}P_i^2-\frac{1}{4}[X_i,X_j]^2),
\end{equation}
where $P_i$ is the momentum conjugate to the $X_i$.

It has been proved in \cite {WLN}
that the Hamiltonian of the supersymmetric
matrix model has a continuous spectrum starting at zero.
This result can be interpreted as a manifestation of the instability
of the supermembranes against deformations into stringlike configurations.
The possible instability of membranes is already evident from the
classical consideration because the potential energy has valleys
through which certain membrane configurations can escape to infinity
without increasing the mass.  For the bosonic membrane this classical
instability is cured by quantum mechanics: the spectrum of the
quantum Hamiltonian is actually discrete \cite {Lus,Sim}.  However,
the theory still become unstable if one  introduces supersymmetry
\cite {WLN}. These effects can be seen in a toy bosonic model
\cite{BMS} with the Hamiltonian \begin{equation} \label {1.5}
H=\frac{1}{2}(p_1^2+p_2^2)+\frac{1}{2}x_1^2 x_2^2.
\end{equation}
The quantum mechanical Hamiltonian (\ref {1.5}) has a discrete
spectrum \cite{Lus,Sim,Med}. However, its supersymmetric
version has a continuous
spectrum \cite{WLN}. One observes that certain properties of the
classical bosonic system are more similar to the
properties of the  quantum supersymmetric system rather then to the quantum
bosonic system.  Supersymmetry saves classics.

The dynamical system with the Hamiltonian (\ref {1.5}) is very
interesting because it exhibits a complicated chaotic behavior
of trajectories.
Classical and quantum chaos in the system (\ref {1.5}) obtained
as the reduction of the Yang-Mills theory
has been considered in  \cite {BMS,CS,Med},see also \cite{Shur,Holger}.
Chaos in the Einstein-Yang-Mills   equations was discussed in
\cite{Bar,Galt}.

The potential in (\ref{1.5}) has zero-directions, the valleys, $x_1=0$ and
$x_2=0$.  In these  directions the particle can move without changing energy
and the presence of these hyperbolic
valleys is an origin of stochastic behavior
of the particle.  A typical trajectory is plotted on Fig.1. We see that a
movement of a particle  is limited by four hyperbolas. The particle
being positioned in one of the valleys
starts to oscillate between two nearest hyperbolas and after a number of
oscillations it changes the valley. So, during the time large enough one can
see the particle in any of four valleys. Therefore we cannot treat the
dynamics perturbatively just making a linearisation around some fixed value
of $x_1$ and $x_2$.
To illustrate this instability we present on Fig.2 a
dependence of $\chi (t)= \frac1t \log \rho (t)/\rho(0)$  on t for some
initial data.  We see that for t large enough  $\chi (t)$ goes to a
fixed value, $\chi\approx 0.85$.
\par
Note, that the addition of fermions leads to an appearing of the interaction
with a fermionic current. The new system also has a tendency to produce
a chaotic behavior (cf. \ref{CGM}).

The system (\ref {1.5})
is, in fact, a particular case of the M(atrix) model (\ref {1.1}).
Let us consider the gauge group $SU(2)$ and take the following ansatz:
\begin{equation} \label {1.6}
X_1=x_1\sigma_1, ~X_2=x_2\sigma_2,~X_3=x_3\sigma_3,
\end{equation}
$$
X_4=...=X_9=0,
$$
where $x_{\alpha}=x_{\alpha}(t)$ are real valued
functions of time, $\alpha=1,2,3$ and $\sigma_{\alpha}$ are the Pauli
matrices. Then the constraint (\ref{1.3}) is
satisfied and eqs. (\ref {1.2}) are reduced to
\begin {equation}\label {1.7}
\ddot{x}_{\alpha}=-2(\sum_{\beta\not=\alpha}x^2_{\beta})x_{\alpha}.
\end{equation}
Eqs (\ref {1.7}) could be obtained from the Hamiltonian
\begin{equation}\label {1.8}
H=\frac{1}{2}(p_1^2+p_2^2+p_3^2)
+(x_1^2x_2^2+x_1^2x_3^2+x_2^2x_3^2).
\end{equation}
The system (\ref{1.5})
is obtained from (\ref{1.8}) by seting $x_3=0$ and rescaling the coupling
constant.
The system (\ref {1.8}) is an example of the dynamical
system which exhibits a chaotic motion.
It has been
analyzed numerically in \cite{CS}. On Fig.3 we show $(x_1,x_2)$
projection of a typical trajectory of the particle and
on Fig.4 we plot the corresponding $\chi(t)$.
During the time large enough one can
see the particle in any of six valleys.
By comparing Fig.2 and Fig.4 we see that in the three-dimensional
case $\chi (t)$ tends to a constant value faster then in the
two-dimensional case.

The ansatz (\ref{1.6})  can also be tested on su(3)
\be
X_i=x_iT^i,\;i=1,..8;\;X_9=0,
\ee
where $T^i$ are the su(3) generators. The Lagrangian equations of
motion put the constraint on $x_i$
\be
\label{constr}
x_4^2+x_5^2-x_6^2-x_7^2=0.
\ee
One of the possible solutions of (\ref{constr}) is $x_4=x_6$,
$x_5=x_7$. It yields the system with six degrees  of freedom with
the Hamiltonian
\be
H=\frac{1}{2}\sum_{i=1}^3p_i^2+\frac{1}{4}(p_4^2+p_5^2)+\frac{1}{2}p_8^2
+\frac{1}{2}\sum_{1\le i<j\le 3}x_i^2 x_j^2 +\frac{1}{4}(x_4^2 +x_5^2)
\sum_{1\le i\le 3} x_i^2$$ $$ +\frac{5}{4}x_4^2x_5^2 +\frac18
(x_4^4+x_5^4) + \frac{3}{4}(x_4^2+x_5^2)x_8^2,
\ee
where $p_i$ is momentum conjugated to $x_i$. The estimation of the Lyapunov
exponent by the computer simulation exhibits the stochastic behaviour of
the system. For the particular case $x_2=x_3=x_5=x_8=0$ the Hamiltonian
has the form
\be
\label{h14}
H=\frac{1}{2}p_1^2+\frac{1}{4}p_4^2
+\frac{1}{2}+\frac{1}{4}x_4^2 x_1^2 +\frac18
x_4^4.
\ee
A typical trajectory for (\ref{h14}) is plotted on Fig. 5 and it can
be seen that they are located in the valley $x_1=0$.  The
corresponding Lyapunov exponent is presented on Fig.6.

Note that the ansatz (\ref{1.6})  deals only with the matrix
diagonal in (space, isotopic) indices and all other degrees of freedom are
frozen. One can think that stochastic behavior is an artifact of this ansatz
and instability will be cured by taking into account the fluctuation of the
frozen degrees of freedom. To analyze this possibility we consider the model
(\ref{1.1})  with only two matrices and relaxed constraints.
A relaxing of the Gauss law means that we consider an interaction
of the "electric" field with some current.
Denote: $X_1 =\phi_1$ and $X_2 =\phi_2$, then the action reads
\be \label{action}
S=\int dt\Tr\left(\frac{1}{2}\dot\phi_1^2+\frac{1}{2}\dot\phi_2^2+
 \frac{\lambda}{2}[\phi_1,\phi_2]^2\right).
\ee
This action describes the dynamics of the classical vacuum moduli space in
the $N=2$ SUSY Yang-Mills theory \cite{SW}.

The dynamics of "fast" modes in (\ref{action}) can be reduced in special cases to the following two-dimensional Hamiltonian systems
\be
H=\frac12 (p_1^2+p_2^2)+\lambda x_1^2x_2^2 +\frac{(m-n)^2}{4(x_1-x_2)^2}
+\frac{(m+n)^2}{4(x_1+x_2)^2},
\label{h1}
\ee
\be
H=\frac12 (p_1^2+p_2^2)+\lambda x_1^2x_2^2 +\frac{m^2}{2x_1^2},
\label{h2}
\ee
where $m$ and $n$ are some real constants (see Appendix, where
$x_1=f$, $x_2=g$). These constants describe a contribution from the
"slow" modes to the Hamiltonian of the "fast" modes. We see that
the "slow" modes are separated but they bring a singular term to the
Hamiltonian of the "fast" modes. Note that an influence of the extra
term $1/2x_1^2$ on the chaotic behavior of a two-dimensional system
has been discussed in the recent paper \cite{CGM}.  Typical
trajectories for the Hamiltonian (\ref{h1}) are plotted on Fig.7 and
for the Hamiltonian (\ref{h2}) on Fig.8.

In these special cases one can recover the constraints by taking $m=0$.
In the case (\ref{h2}) one comes
back to the ansatz (\ref{1.8}) with $x_3=0$.
But in the case (\ref{h1}) we get an extra repulsive potential. This
potential describes a repulsive from walls
located along two diagonals $x_1=\pm
x_2$ in the $x_1,x_2$ plane.
A movement of a particle is limited by equipotential lines
$$
\lambda x_1^2x_2^2 +\frac{n^2(x_{1}^{2}+x_{2}^{2})}{2(x_1^2-x_2^2)^{2}}
=const,
$$
and the particle oscillates in one of four valleys (it chooses one of them
according to the initial data). In each of four allowed domains
the potential still has a direction of the instability.
Therefore,
the presence of the constraint does not affect the chaotic character of
trajectories.

Different approaches have been developed to address the question what
is the way in which classical chaos manifests itself in the
properties of the corresponding quantum system \cite{Zas,Gut},
\cite{BMS}-\cite{CS}, \cite{GVZ}-\cite{CGM}. The simplest manifestation of
classical chaos for a quantum system is the nature of spectral fluctuations
of the energy levels. It was suggested  that for chaotic systems
the statistical properties of the spectrum should be that of the random
matrix theory with the Wigner-Dyson distribution
\begin{equation} \label{1.10}
P(E|\Delta E)=A|\Delta E|^{\alpha}\exp[-B(\Delta E)^2],
\end{equation}
where
$\alpha >0$, whereas the quantum version of a classically integrable system
is described by the Poisson distribution
\begin{equation} \label {1.11}
P(E|\Delta E)=a\exp[-b|\Delta E|]
\end{equation}

The crucial difference between  (\ref {1.10}) and  (\ref {1.11})
is the behavior for $\Delta E \to 0$. For the Wigner-Dyson distribution
(\ref {1.10}) one has $P(E|\Delta E) \to 0$ if  $\Delta E \to 0$, i.e.
the density of the energy levels at the small scale vanishes.
This is interpreted as the repulsive feature of the energy eigenvalues
for quantum chaotic system.  One can try to relate this feature
with the holographic feature of M(atrix) theory.
The holographic principle follows from the
general consideration involving the Bekenstein-'t Hooft bound
on the entropy of a spatial region. A
holographic theory contains only degrees of freedom which carry the smallest
unit of longitudinal momentum \cite {BFSS}. The transverse density
of partons is bounded to about one parton per transverse Planck area, in this
sense the holographic theory is repulsive. The partons form a kind of
incompressive fluid. The repulsive feature of the holographic
theory seems to be related with the repulsive feature of the distribution of
the energy levels for quantum chaotic system.  Quantum chaos is a source of
the holographic feature of M-theory.  A connection between the large $N$
limit, random matrix theory and entropy of black holes is discussed in \cite
{Vol}-\cite {DVV}.

In conclusion, we have shown that M(atrix) theory  exhibits
the classical chaotic motion and we have argued that
quantum chaos is a source of the holographic
feature. The statistical properties of the spectrum
and other features of quantum chaos in M-theory require a further
investigation.

$$~$$
{\bf ACKNOWLEDGMENT}
$$~$$
The authors are grateful to B.V.Medvedev for stimulating discussions.
I.A., P.M. and O.R.  are supported
in part by  RFFI grant 96-01-00608.
I.V. is supported  in part by RFFI grant 96-01-00312.

$$~$$

\appendix
\section*{Appendix}
\renewcommand{\theequation}{A.\arabic{equation}}
\setcounter{equation}{0}

Here we shall explore the classical dynamics of the
system with the action (\ref{action})
and deduce the Hamiltonians (\ref{h1}) and (\ref{h2}).

The Lagrangian admits two global continuous symmetries. Since the fields $\phi_1$ and
$\phi_2$ are the elements of the su(2) algebra the action
(\ref{action}) is invariant under
\be
\phi_1^{'}=U^{+}\phi_1 U,\qquad \phi_1^{'}=U^{+}\phi_1 U,
\ee
$U\in SU(2)$. This symmetry yields the  conservation
of the "angular momentum"
\be
\label{M}
M_0 =[\phi_1,\dot\phi_1]+[\phi_2,\dot\phi_2].
\ee
Another symmetry corresponds to U(1) transformations which mix the fields
$$
\phi_1^{'}=\cos\theta\,\phi_1-\sin\theta\,\phi_2,
$$
\be
\phi_2^{'}=\sin\theta\,\phi_1+\cos\theta\,\phi_2.
\ee
It gives one more first integral
\be
N=\Tr(\phi_1\dot\phi_2-\dot\phi_1\phi_2).
\ee
The equations of motion for the Lagrangian (\ref{action})
read
\be
\stackrel{\cdot\cdot}{\phi_1}=2\lambda [\phi_2,[\phi_1,\phi_2]],\qquad
\stackrel{\cdot\cdot}{\phi_2}=2\lambda [\phi_1,[\phi_2,\phi_1]].
\ee

It is convenient to parametrize $\phi_1$ and $\phi_2$ as follows:
$$
\phi_1(t)=\sqrt{2}U^{+}(t)\left(\frac{\sigma_3}{2}f(t)\cos\theta(t)-
\frac{\sigma_2}{2}g(t)\sin\theta(t)\right)U(t),
$$
\be
\phi_2(t)=\sqrt{2}U^{+}(t)\left(\frac{\sigma_3}{2}f(t)\sin\theta(t)+
\frac{\sigma_2}{2}g(t)\cos\theta(t)\right)U(t),
\label{uu}
\ee
where $f(t),\;g(t),\;\theta(t)$ are real functions and $U(t)$ is a
SU(2) group element.

Let us analyze the consistency of the parametrization (\ref{uu}).
The variables $\phi_1$ and $\phi_2$ could be treated as vectors
in the internal isotopic space.
At any time they can be rotated to belong to some coordinate plane, say
(2,3),  by using an $U(t)\in  SU(2)$:  \be
\phi_1=(0,\,\psi_1,\,\psi_2),\qquad\phi_2=(0,\,\psi_3,\,\psi_4),
\label{p}
\ee
that fixes $U(t)$ up to rotation around the 1-axis.
This rotation could be used to impose the following constraint:
\be
\psi_1\psi_2+\psi_3\psi_4=0.
\label{co}
\ee
The rotation angle to fulfill (\ref{co}) is
$$
\tan 2\chi=-\frac{2(\psi_1\psi_2+\psi_3\psi_4)}{\psi_1^2-\psi_2^2+\psi_3^2-
\psi_4^2}.
$$
Two vectors in the plain obeying (\ref{co}) could be parametrized by three
variables $f,\,g,\,\theta$. Namely :
$$f=\frac{\psi_4}{\psi_1}\sqrt{\psi_1^2+\psi_2^2},~~
g=-\sqrt{\psi_1^2+\psi_3^2},~~
\tan\theta=-\frac{\psi_1}{\psi_3}$$

In this coordinate system the Lagrangian acquires the form
$$
L= \frac{1}{2} (\dot f^2 +\dot g^2 +
(f^2 +g^2)\dot \theta^2 + (f^2 +g^2)\Tr \dot U \dot U^+)
-2i\Tr \dot U U^+
\sigma_1 fg\dot \theta
$$
\be
\label{lagr_U}
+\frac{1}{2}f^2
\Tr \dot U U^+ \sigma_3 \dot U U^+\sigma_3+
\frac{1}{2}g^2 \Tr \dot U U^+
\sigma_2 \dot U U^+\sigma_2 - \lambda f^2g^2 .
\ee

Using "fast-slow" modes terminology we can say that $f$ and $g$
are the "fast" modes and $\theta$ and $U$ describe the "slow" modes.

Note that $\theta$ is a cyclic coordinate and
the corresponding first integral is just $N=n$ with
\be
N=(f^2 +g^2)\dot \theta +2fg l_1,
\label{N}
\ee
here we denote: $\dot U U^+ =l=\frac i2 \sigma_j l_j$.

The equations of motion for $f$ and $g$ in this parametrization read:
\be
\stackrel{\cdot\cdot}{f}=f\dot\theta^2+f(l_1^2 +l_2^2)
+2g\dot\theta l_1 -2\lambda fg^2,
\label{f}
\ee
\be
\stackrel{\cdot\cdot}{g}=g\dot\theta^2+g(l_1^2 +l_3^2)+
 2f\dot\theta l_1 -2\lambda f^2g,
\label{g}
\ee
The remaining three equations of motion are nothing but the angular
momentum conservation law $\dot M_0 =0$. It is convenient to
introduce $M$ - the angular momentum in the moving frame $M=UM_0U^+
$ with the covariant conservation law
\be
\dot M +[M,l]=0.
\label{a}
\ee
The explicit expression for $M$ reads
\be
M= i\sigma_1 \left( (f^2 +g^2)l_1 +2fg\dot \theta\right) +
i\sigma_2 (f^2l_2)+i\sigma_3(g^2l_3).
\label{b}
\ee
The Lagrangian can be expressed in terms of $M$ and $N$ as follows
\be
L=\frac12(\dot f^2 +\dot g^2) -\lambda f^2g^2 +
\frac12 (\dot \theta N - \Tr l M).
\label{la}
\ee

There is a wide class of solutions with $[M,l]=0$, i.e. where $M$ is
conserved. It follows from eq.(\ref{b}) that in this case only one component
of $l$ can be non-zero.

In the first case: $l_2=l_3=0$ eqs. (\ref{N}), (\ref{a}) and (\ref{b}) give
\begin{eqnarray}
2fgl_1 +(f^2 +g^2)\dot \theta&=&n\nonumber \\
(f^2 +g^2)l_1 +2fg\dot \theta &=& m. \nonumber
\end{eqnarray}
The matter of simple calculation to verify that the dynamics of the $(f,g)$
system (eqs. (\ref{f}), (\ref{g})) is governed by the "effective" potential
\be
\label{cross}
V=\lambda f^2g^2+\frac{(m-n)^2}{4(g-f)^2}+\frac{(m+n)^2}{4(g+f)^2}.
\ee
\be
\stackrel{\cdot\cdot}{f} =-\frac{\partial V}{\partial f}, ~~~~~
\stackrel{\cdot\cdot}{g} =-\frac{\partial V}{\partial g}.
\ee
The second case: $l_1=l_3=0$, gives
$$
M_1=2fg\dot \theta
$$
and to keep $ M$ parallel to $ l$ one has to put $\dot \theta =0$.
In this case the effective potential is
\be
\label{line}
V=\lambda f^2g^2+\frac{m^2}{2f^2},
\ee
where $m=f^2l_2$ is the first integral.

The third case $l_1=l_2=0$ is quite similar to the second one with the
exchange 2$\to \hspace{-4mm}\gets$3.

\newpage \par
 \epsfig{file=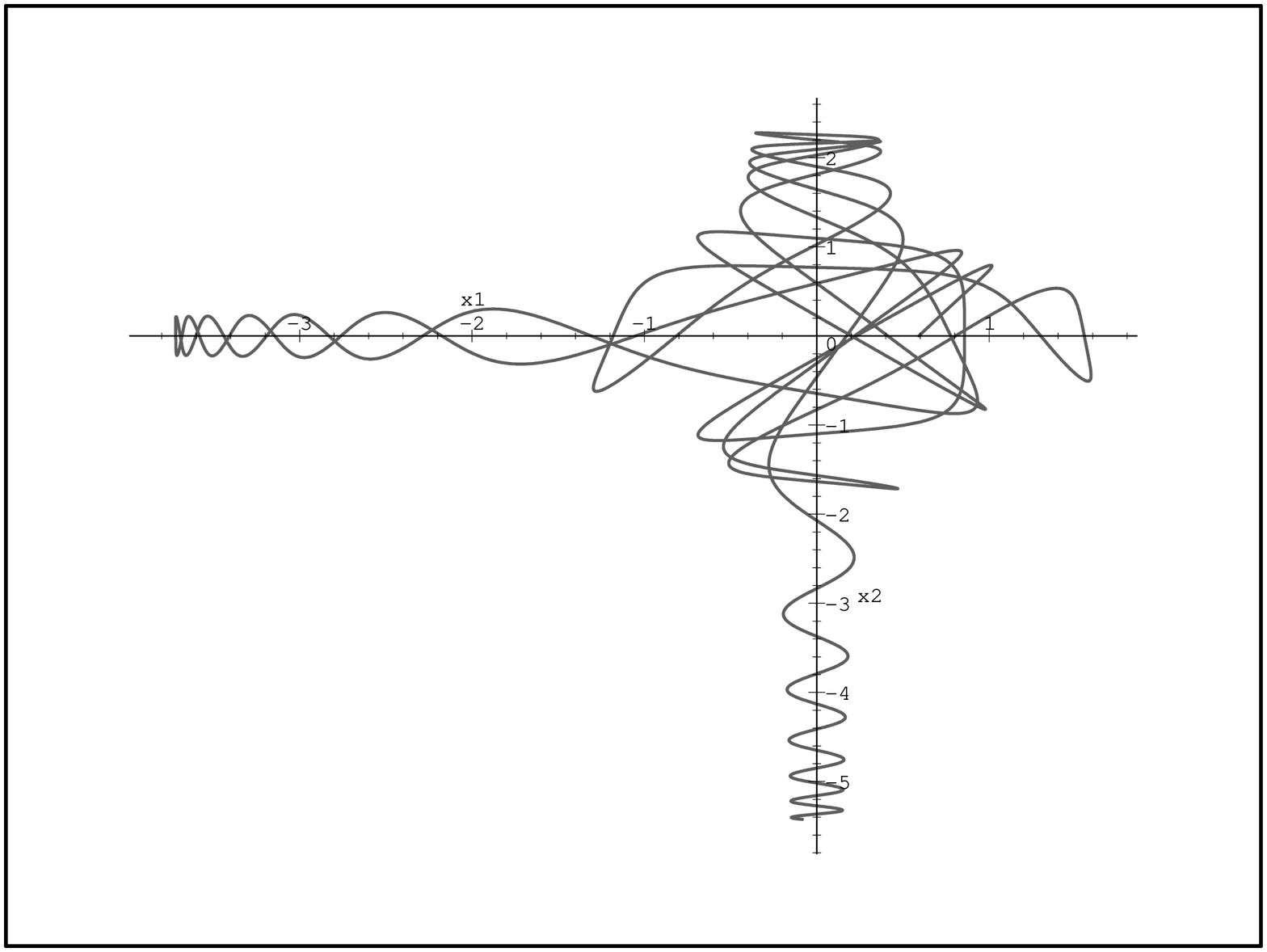,
   width=350pt,
   angle=-90
 }
\begin{center}
Fig. 1. Typical trajectory of the two-dimensional system with Hamiltonian
(\ref{1.5})
\end{center}
\par
 \epsfig{file=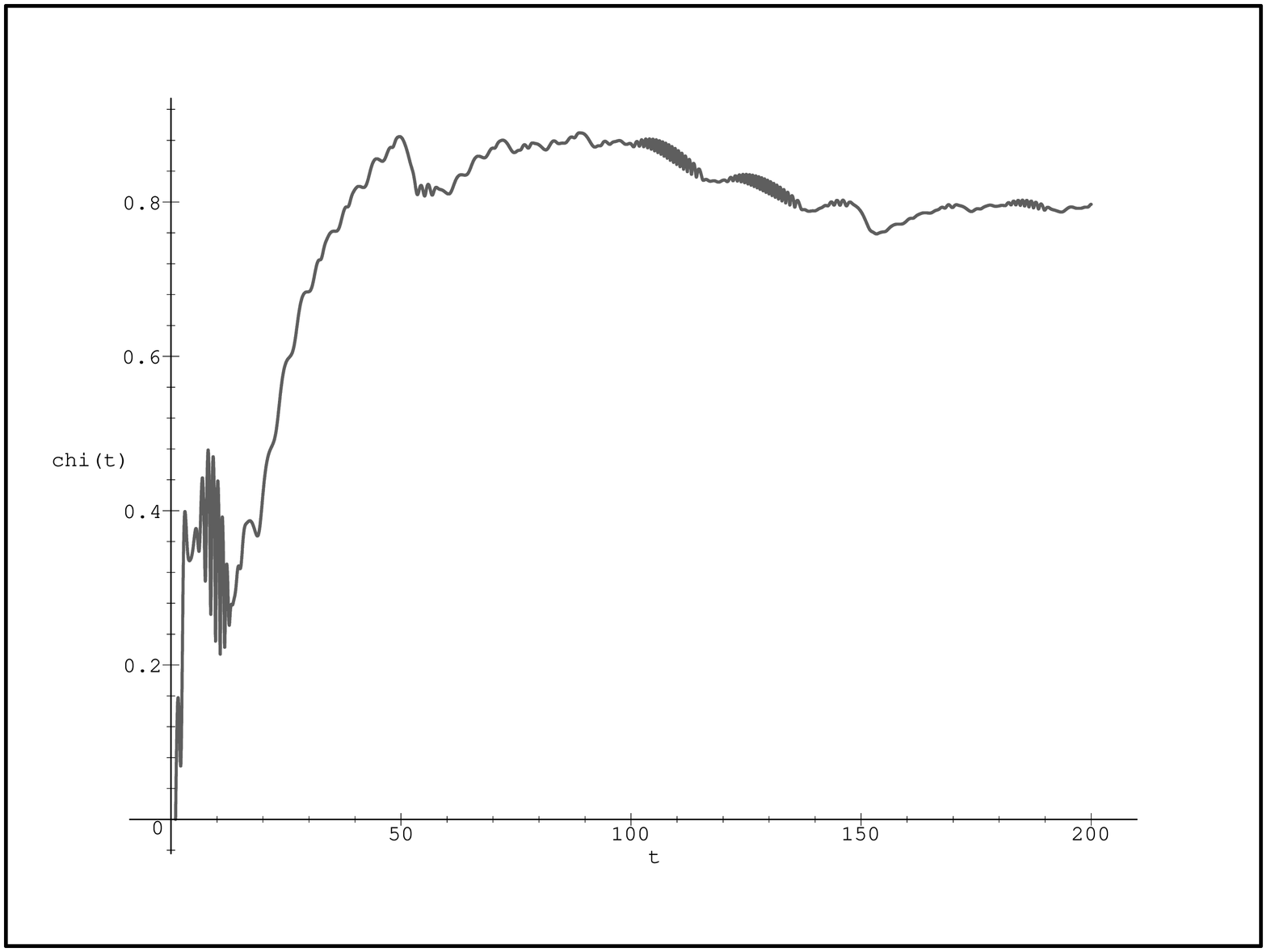,
   width=350pt,
   angle=-90
 }
\begin{center}
Fig. 2. Lyapunov exponent of the two-dimensional system with Hamiltonian
(\ref{1.5})
\end{center}
\par
 \epsfig{file=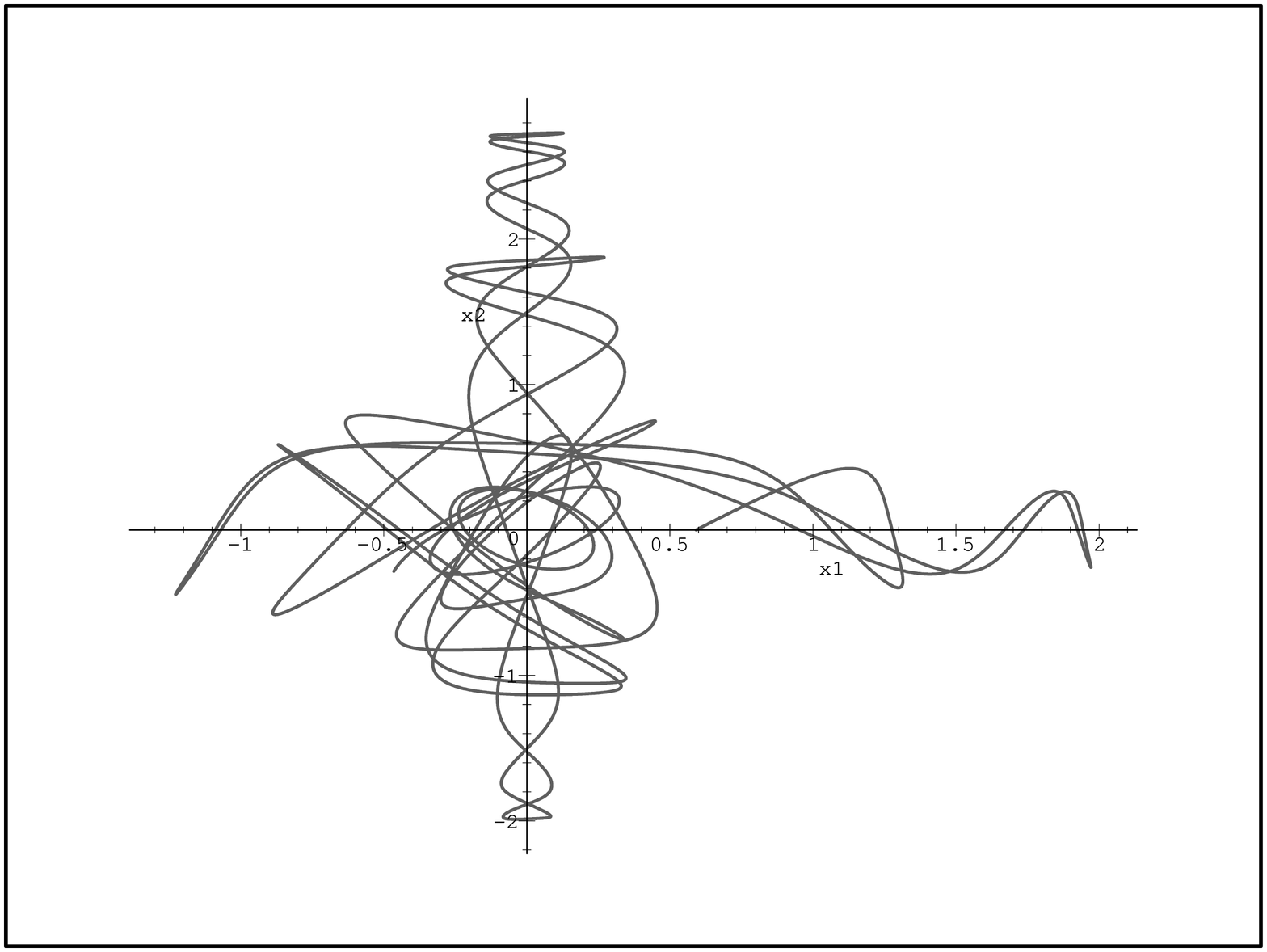,
   width=350pt,
   angle=-90
 }
\begin{center}
Fig. 3. Typical trajectory of the three-dimensional system with Hamiltonian
(\ref{1.8})
\end{center}
\par
 \epsfig{file=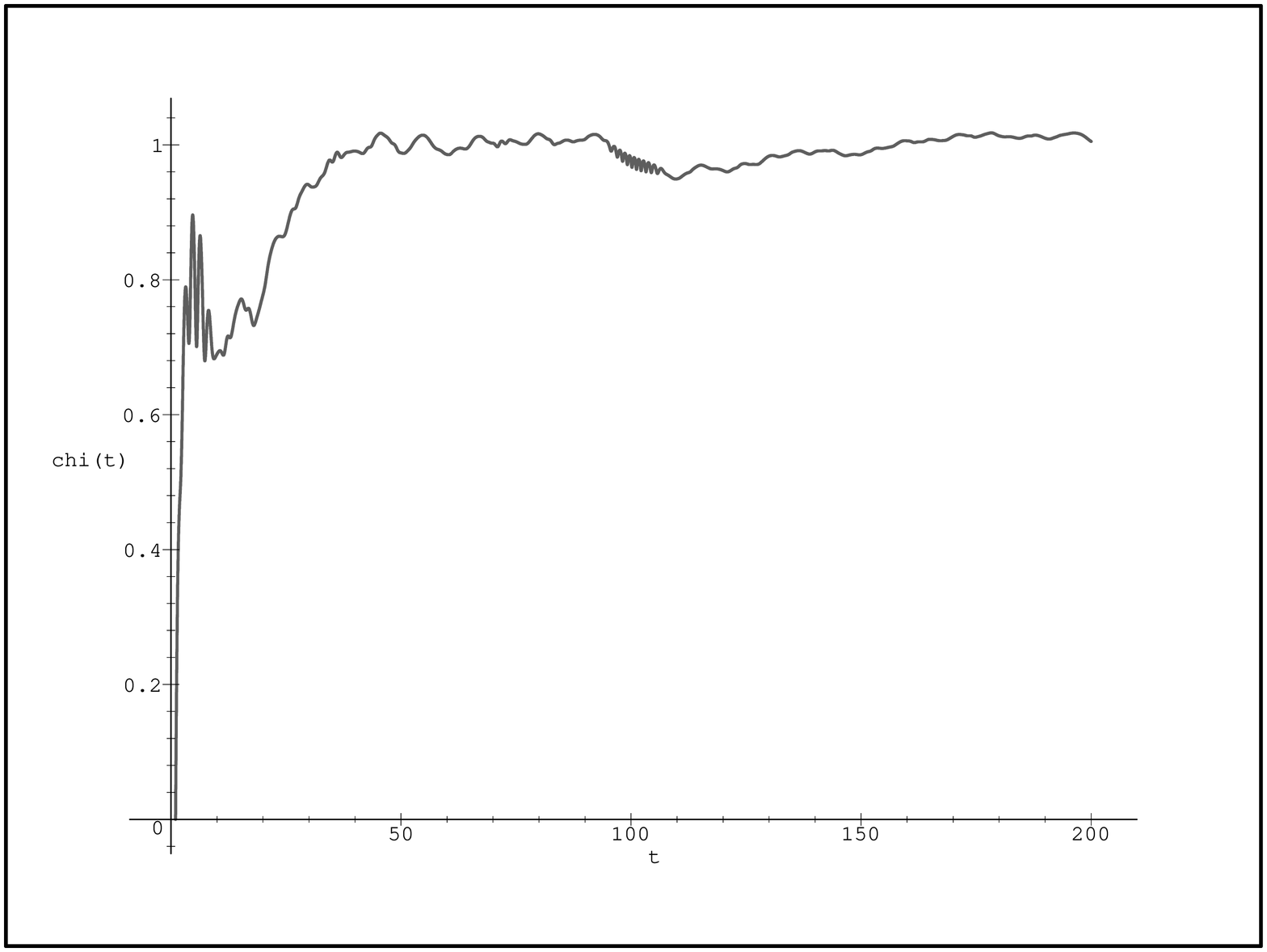,
   width=350pt,
   angle=-90
 }
\begin{center}
Fig. 4. Lyapunov exponent of the three-dimensional system with Hamiltonian
(\ref{1.8})
\end{center}
\par
 \epsfig{file=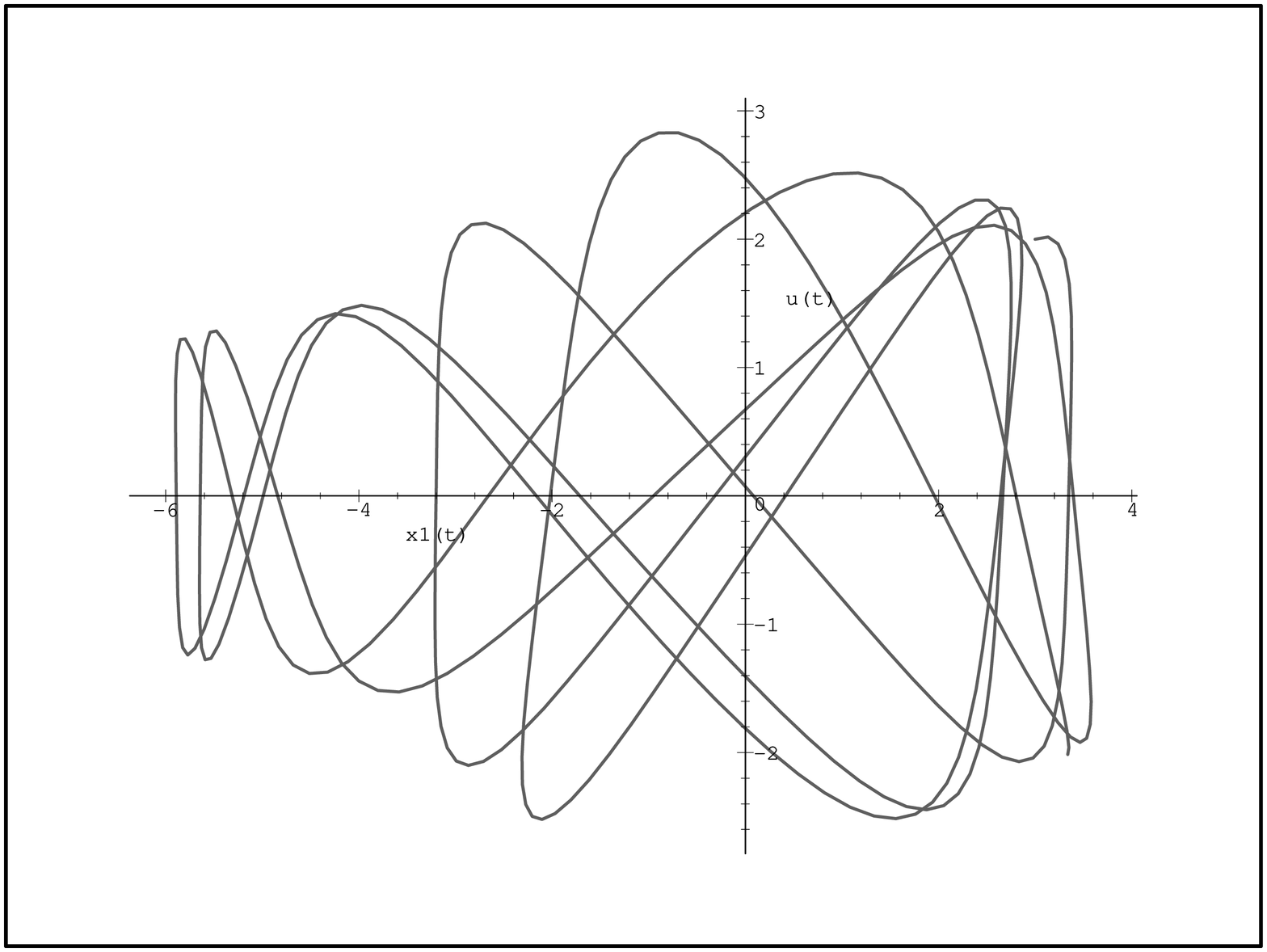,
   width=350pt,
   angle=-90
 }
\begin{center}
Fig. 5. Typical trajectory of the two-dimensional system with Hamiltonian
(\ref{h14}).
 \end{center}
\par
 \epsfig{file=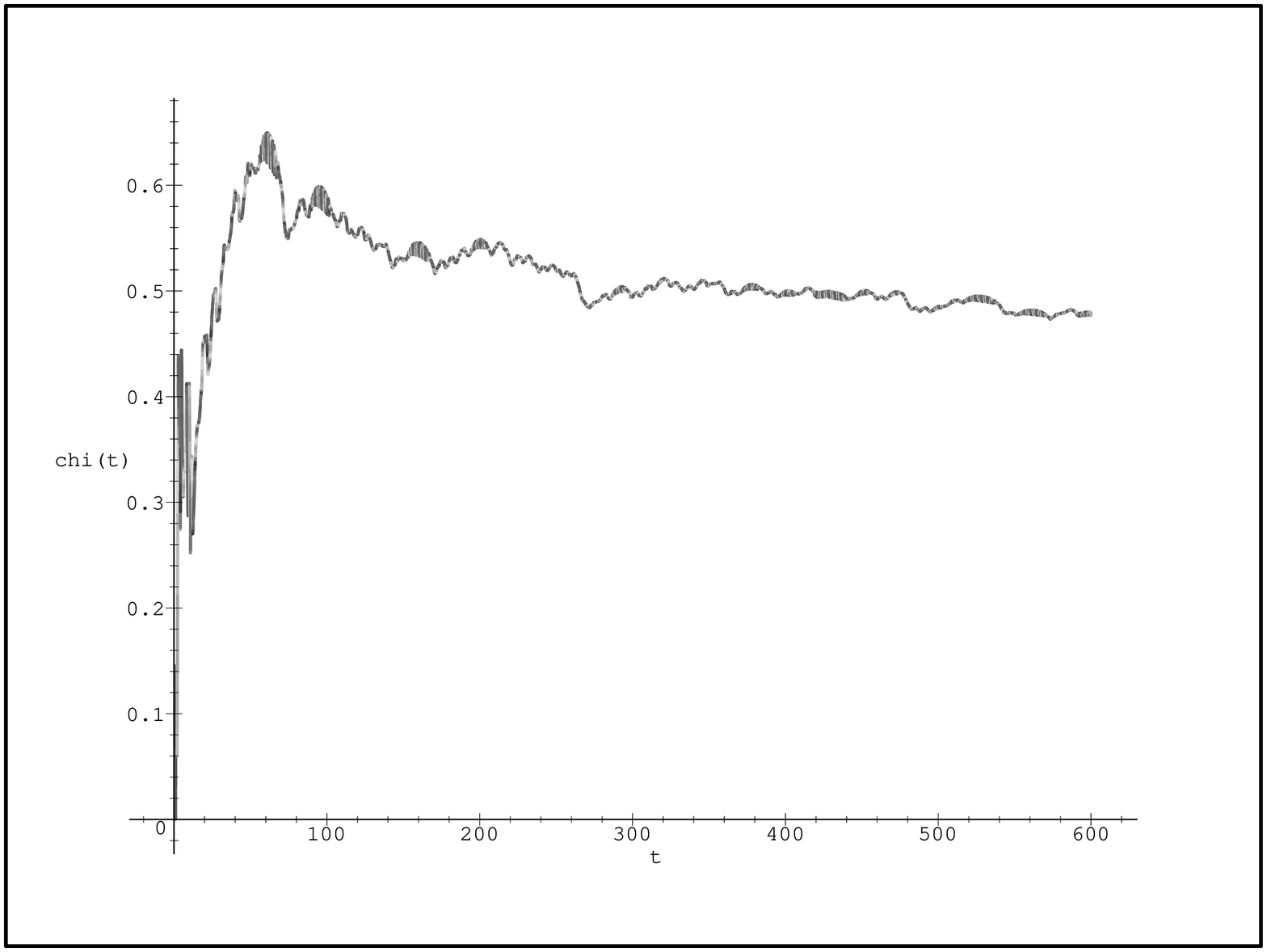,
   width=350pt,
   angle=-90
 }
\begin{center}
Fig. 6. The example of the Lyapunov exponent for the two-dimensional
system, obtained in the su(3) case.
 \end{center}
\par
 \epsfig{file=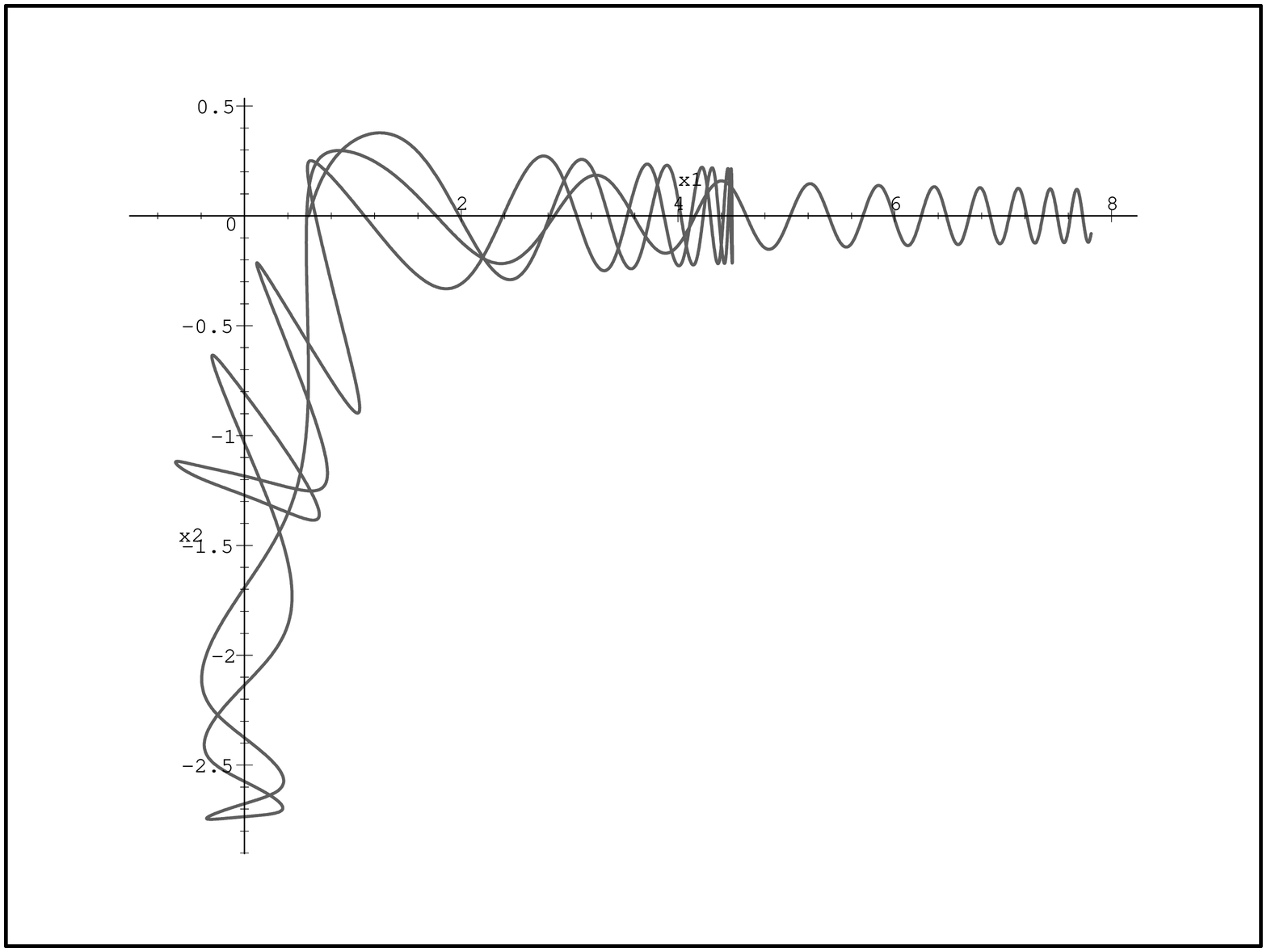,
   width=350pt,
   angle=-90
 }
\begin{center}
Fig. 7. Typical trajectory of the  system with Hamiltonian
(\ref{h1}), $m=-n$ \end{center} \par
\epsfig{file=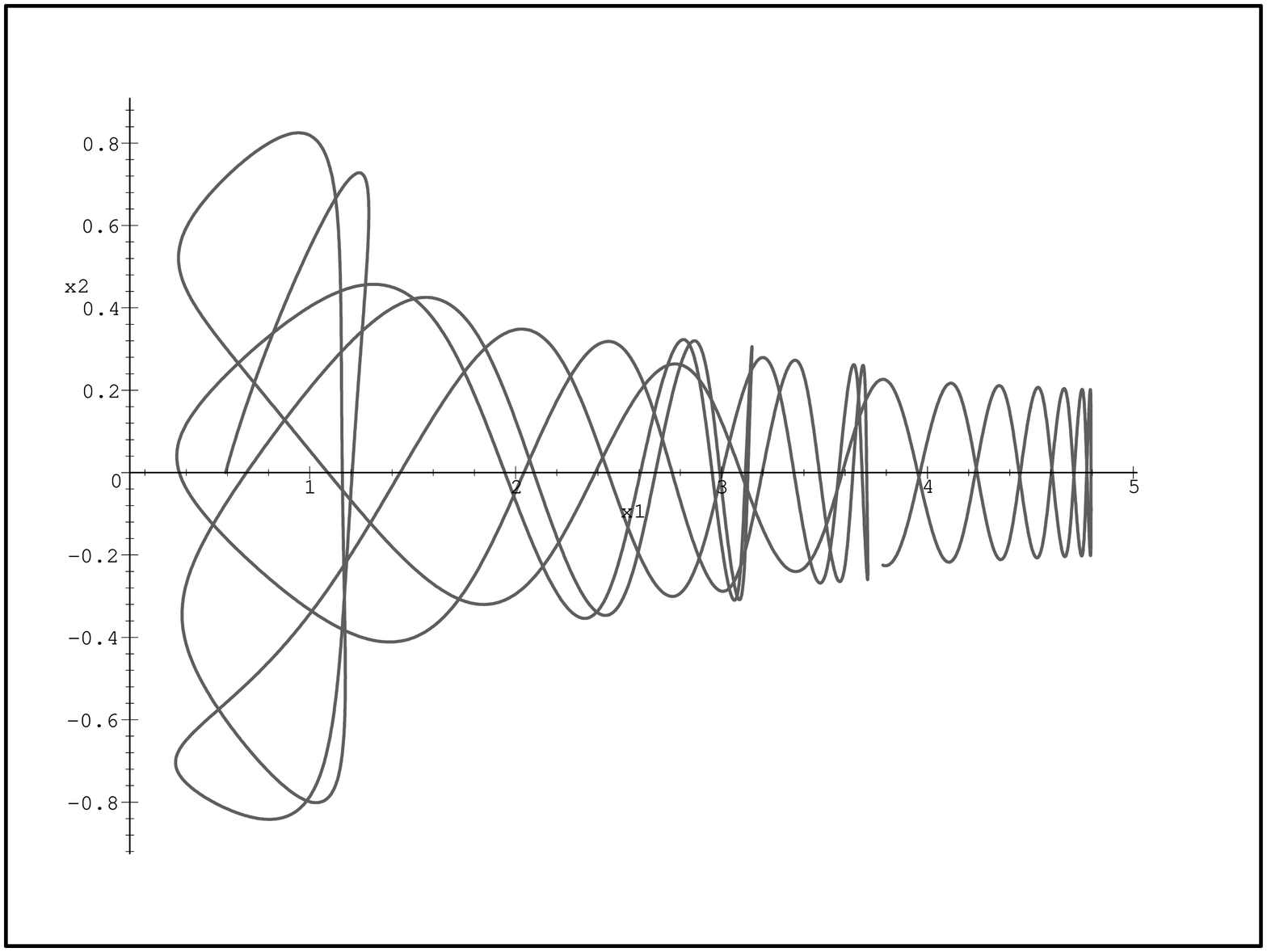,
   width=350pt,
   angle=-90
 }
\begin{center}
Fig. 8. Typical trajectory of the  system with Hamiltonian
(\ref{h2}) \end{center}
\par


{\small
\begin{thebibliography}{99}
\bibitem{HT} C. M. Hull and P. K. Townsend, {\it Nucl. Phys.} B438(1995)109
\bibitem{Wit} E. Witten, {\it Nucl. Phys.} B443(1995)85
\bibitem{BFSS} T. Banks, W. Fischler, S. H. Shenker and L. Susskind,
{\it M Theory as a Matrix Model: a Conjecture}, {\it Phys. Rev.}
D55(1997)5112, hep-th/9610043
\bibitem{WHN} B. de Wit, J. Hoppe and H. Nicolai, {\it Nucl. Phys.}
B305 (1988)545
\bibitem{WLN} B. de Wit,  M.Luscher and H. Nicolai, {\it Nucl. Phys.}
B 320 (1989) 135
\bibitem{FH} J. Fr$\ddot{o}$hlich and J. Hoppe ,
{\it On Zero-Mass Ground States
in Super-Membrane Matrix Models}, ITH-TH/96-53
\bibitem{DFS} U. H. Danielsson, G. Ferretti and B. Sundborg, hep-th/9603081
\bibitem{KP} D. Kabat and P. Pouliot, hep-th/9603127
\bibitem{Sus} L. Susskind, hep-th/9704080
\bibitem{AA} D. V. Anosov and V. I. Arnold  (eds),
{\it Dynamical Systems}, VINITI, Moscow, 1985
\bibitem{Sin} Ya. G. Sinai, {\it Introduction to Ergodic Theory},
Fasis, Moscow, 1996
\bibitem{ER}  J.-P. Eckmann and D. Ruelle, {\it Rev. Mod. Phys.}
57(1987)617
\bibitem{Zas} G. M. Zaslavskii, {\it Stochasticity  of Dynamical
Systems,} Nauka, Moscow, 1984
\bibitem{Gut} M. C. Gutzwiller, {\it Chaos in Classical and Quantum
Mechanics}, Springer, Berlin, 1990
\bibitem{Ohya} M. Ohya, {\it Complexities and their Applications
to Characterization of Chaos}, to appear in Intern. J. of Theor.
Physics, 1997
\bibitem{Lus} M. L$\ddot{u}$scher, {\it Nucl. Phys.} B219(1983)233
\bibitem{Sim} B. Simon {\it Ann. Phys.} 146(1983)209
\bibitem{BMS} G. Z. Baseyan, S. G. Matinyan and G. K. Savvidi,
{\it JETP Lett.} 29(1979)585
\bibitem{Med} B. V. Medvedev, {\it Teor. Mat. Phys.} 60(1984)224;
109(1996)406
\bibitem{CS} B. V. Chirikov and D. L. Shepelyanskii, {\it JETP Lett.}
34(1981)164
\bibitem{Shur}
   E.S. Nikolaevsky and L.N. Shchur,
   JETP LEtt., {\bf 36} (1982) 218-220;
   JETP {\bf 58} (1983) 1
\bibitem{Bar} J. D. Barrow and J. Levin,  gr-qc/9706065;
J. D. Barrow, M. P. Dabrowski, hep-th/9711041
\bibitem{Galt} D. V. Gal'tsov and M. S. Volkov, {\it Phys. Lett.} B256 (1991)
17
\bibitem{SW} N. Seiberg and E. Witten, Nucl. Phys.  B426
(1994) 19; B431 (1994) 484.  \bibitem{GVZ} M.-J. Giannoni, A. Voros and J.
Zinn-Justin (eds.), {\it Chaos and Quantum Physics}, North-Holland,
Amsterdam, 1991 \bibitem{Casati} G. Casati and B. V. Chirikov (eds.), {\it
Quantum Chaos: between order and disorder}, Cambridge Univ. Press, Cambridge,
1995 \bibitem{Nak} K. Nakamura, {\it Quantum Chaos}, Cambridge University
Press, Cambridge, 1995
\bibitem{KO} T. Kawabe and S. Ohta, {\it Phys. Rev.} D44(1991)1274
\bibitem{BGC}  O. Bogidas, M.-J. Giannoni and C. Schmit,
{\it Phys. Rev. Lett.} 52(1984)1
\bibitem{JPC} G. Jona-Lasinio, C. Presilla and F. Capasso,
{\it Phys. Rev. Lett.} 68(1992)2269
\bibitem{Sre} M. Srednicki, cond-mat/9605127
\bibitem{Holger} H. B. Nielsen, H. H. Rugh and S. E. Rugh,
chao-dyn/9605013, hep-th/9611128
\bibitem{ AL} B. L. Altshuler and L. S. Levitov, cond-mat/9704122
\bibitem{LU} L. Salasnich, {\it Mod. Phys. Lett.} A12 (1997)
1473-1480, quant-ph/9706025
\bibitem{CGM} L. Casetti, R. Gatto  and M.
Modugno, hep-th/9707054
\bibitem{Vol} I. V. Volovich, {\it D-branes, Black Holes and
$SU(\infty)$ Gauge Theory}, hep-th/9608137
\bibitem{IKKT} N.  Ishibashi, H. Kawai, Y. Kitazawa and A. Tsuchiya, {\it A
Large-N Reduced Model as Superstring}, hep-th/9612115 \bibitem{DVV} R.
Dijkgraaf, E. Verlinde and H. Verlinde, {\it 5D Black Holes and Matrix
Strings}, hep-th/9704018 \end{thebibliography} }
\end{document}